\documentclass[twocolumn,showpacs,preprintnumbers,amsmath,amssymb,prl]{revtex4}

\usepackage{graphicx}
\usepackage{dcolumn}
\usepackage{bm}

\usepackage{latexsym}

\begin{document}

\preprint{}

\title{Topological aspects of quantum spin Hall effect in graphene: \\
Z$_2$ topological order and spin Chern number}

\author{Takahiro Fukui}
 \affiliation{Department of Mathematical Sciences, Ibaraki University, Mito
310-8512, Japan}
\author{Yasuhiro Hatsugai}
\affiliation{Department of Applied Physics, University of Tokyo, Hongo,
Tokyo 113-8656, Japan}

\date{July 19, 2006}

\begin{abstract}
For generic time-reversal invariant systems with spin-orbit couplings,
we clarify a close relationship between the Z$_2$ topological order 
and the spin Chern number proposed by Kane and Mele and
by Sheng {\it et al.}, respectively, in the quantum spin Hall effect.
It turns out that a global gauge transformation
connects different spin Chern numbers (even integers) modulo 4,
which implies that the spin Chern number and
the Z$_2$ topological order yield the same classification.
We present a method of computing spin Chern numbers and 
demonstrate it in single and double plane of graphene. 
\end{abstract}

\maketitle

Topological orders \cite{Wen89,Hat04} play a crucial role
in the classification of various phases in low dimensional systems.
The integer quantum Hall effect (IQHE)
is one of the most typical examples
\cite{TKNN82,Koh85}, in which the quantized Hall conductance is given by 
a topological invariant, Chern number, due to the Berry phase \cite{Ber84}
induced in the Brillouin zone \cite{Sim83}.
Such a topological feature should be more fundamental, since it has a close 
relationship with the parity anomaly of Dirac fermions 
\cite{Sem84,Ish84,Hal88}.

Recently, the spin Hall effect \cite{MNZ03,Sin04,KMGA04,WKSJ04}
has been attracting much current interest
as a new device of so-called spintronics.
In particular, 
Kane and Mele \cite{KanMel05a,KanMel05b} 
have found a new class of insulator showing
the quantum spin Hall (QSH) effect 
\cite{BerZha05,QWZ05,SSTH05}
which should be realized in 
graphene with spin-orbit couplings.
They have pointed out \cite{KanMel05b} that 
the QSH  state can be specified by a Z$_2$ topological order
which is inherent in time-reversal (${\cal T}$) invariant systems. 
This study is of fundamental importance, since the Z$_2$ order 
is involved with the Z$_2$ anomaly of Majorana fermions
\cite{AtiSin,Wit84}. 

On the other hand, Sheng {\it et al.} \cite{SWSH06} have recently
computed the spin Hall conductance 
by imposing spin-dependent twisted boundary condition,
generalizing the idea of Niu {\it et al.} \cite{NTW85,Hat04}. 
They have shown that it is given by a
Chern number which is referred to as a spin Chern number below.
This seems very natural, 
since the QSH effect is a spin-related version of IQHE.
The spin Chern number for graphene computed by Sheng {\it et al.} 
indeed has a good correspondence with the 
classification by Z$_2$.

However, the Chern number is specified by the set of
integers Z, not by Z$_2$. 
Although 
the studies 
by Sheng {\it et al. } \cite{SWSH06}
suggest a close relationship between two topological orders,
natural questions arise: 
How does the concept Z$_2$ enter into the classification by Chern numbers,
or otherwise, does the spin Chern number carry additional information?

In this letter, we clarify the relationship for generic ${\cal T}$
invariant systems. We show that while two sectors 
in the Z$_2$ classification are separated 
by topological changes due to {\it bulk gap-closing phenomena}, 
each of these sectors is further divided 
into many sectors by {\it boundary-induced topological changes} 
in the spin Chern number classification.
The latter is an artifact which is due to 
calculations in finite size systems with broken translational invariance.
Therefore, 
the different spin Chern numbers in each sector of Z$_2$
describe the same topologically ordered states
of the bulk. 

Consider generic electron systems on a lattice 
with ${\cal T}$ symmetry, described by the Hamiltonian $H$. 
Denote the electron creation operator at $j$th site as
$c^\dagger_j=(c^\dagger_{\uparrow j}, c^\dagger_{\downarrow j})$.
Then, ${\cal T}$ transformation is defined by
$c_j\rightarrow {\cal T}c_j$ with ${\cal T}\equiv i\sigma^2 K$,
where the Pauli matrix $\sigma^2$ operates the spin space 
and $K$ stands for the 
complex conjugation operator. 
Let ${\cal H}(k)$ be the Fourier-transformed Hamiltonian defined by 
$H=\sum_k c^\dagger(k){\cal H}(k)c(k)$ and let $|n(k)\rangle$ be
an eigenstate of ${\cal H}(k)$. 
Assume that the ground state is composed of an $M$-dimensional multiplet of 
degenerate single particle states
which is a generalized non-interacting Fermi sea \cite{Hat04}.
Kane and Mele \cite{KanMel05b} have found that 
the ${\cal T}$ invariant systems have two kinds of important states
belonging to ``even'' subspace and ``odd'' subspace:
The states in the even subspace have the property that $|n(k)\rangle$
and ${\cal T}|n(k)\rangle$ are identical, which occurs when
${\cal T}{\cal H}(k){\cal T}^{-1}={\cal H}(k)$. 
By definition, 
the states at $k=0$ always belong to this subspace.
The odd subspace has the property that the multiplet $|n(k)\rangle$
are orthogonal to the multiplet ${\cal T}|n(k)\rangle$.
These special subspaces can be detected by 
the pfaffian 
$p_{\rm KM}(k)\equiv \mbox{pf}~\langle n(k)|{\cal T}|m(k)\rangle$.
Namely, $p_{\rm KM}(k)=1$ in the even subspace and 0 in the odd subspace. 
Kane and Mele have claimed that 
the number of zeros of $p_{\rm KM}(k)$ 
which always appear as $\cal T$ pairs $\pm k^*$  
with opposite vorticities
is a topological invariant for ${\cal T}$ invariant systems.
Specifically, if the number of zeros in half the 
Brillouin zone is 1 (0) mod 2, the ground state is in the 
QSH (insulating) phase.

We now turn to the spin Chern number 
proposed by Sheng {\it et al.} \cite{SWSH06}.
According to their formulation,  we impose  spin-dependent 
(-independent) twisted boundary condition along $1(2)$-direction
\begin{alignat}{1}
&
c_{j+L_1\hat 1}=e^{i\theta_1 \sigma^3}c_j, \quad
c_{j+L_2\hat 2}=e^{i\theta_2}c_j ,
\label{TwiBouCon}
\end{alignat}
where a set of integers $j\equiv(j_1,j_2)$ specifies the site and
$\hat 1$ and $\hat 2$ stand for the unit vectors in $1$- and
$2$-directions, respectively.
Let $ {\cal H}(\theta)$ denote the twisted Hamiltonian 
and let $|n(\theta)\rangle$ be corresponding eigenstate. 
It follows from Eq. (\ref{TwiBouCon}) that ${\cal T}$ transformation
induces 
${\cal T}{\cal H}(\theta_1,\theta_2){\cal T}^{-1}
={\cal H}(\theta_1,-\theta_2)$ 
and therefore we can always choose
$|n(\theta_1,-\theta_2)\rangle={\cal T}|n(\theta_1,\theta_2)\rangle$
except for $\theta_2=0$. The states on the {\it line} $\theta_2=0$ belong
to the even subspace.
Below, 
the torus spanned by $\theta$ is referred to as twist space.
The boundary condition (\ref{TwiBouCon}) enables us to define the spin Chern number, 
but the cost we have to pay is the broken translational invariance.
In other words, we slightly break ${\cal T}$ invariance at the boundary,
which leads to nontrivial spin Chern numbers. The average over the twist
angles recovers ${\cal T}$ invariance.

Let us define a 
pfaffian for the present twisted system as a function of the twist angles: 
\begin{alignat}{1}
p(\theta)\equiv \mbox{pf}~\langle n(\theta)|{\cal T}|m(\theta)\rangle,
\quad n,m=1,\cdots,M ,
\label{Pfa}
\end{alignat}
where $M$ is a number of one particle states below the Fermi energy with the
gap opening condition.
Since the line $\theta_2=0$ belongs to the even subspace,
the zeros of the pfaffian in the $\theta_2>0$ or $<0$ twist space
can move only among the same half twist space keeping their vorticities,
and never cross the $\theta_2=0$ line. 
Therefore, one ${\cal T}$ pair of zeros are never annihilated, like those
of the KM pfaffian $p_{\rm KM}(k)$.
Furthermore,
the two zeros in the same half twist space are 
never annihilated unless they have the opposite vorticity.
This is a crucial difference between the KM pfaffian and the twist pfaffian.
The number of zeros in the twist pfaffian 
should be classified by even integers
(half of them, by Z).

Are these pfaffians topologically different quantities?
The answer is no.
To show this, let us consider a global gauge transformation 
$c_j\rightarrow g(\varphi) c_j$, where
\begin{alignat}{1}
g(\varphi)\equiv
e^{i\sigma^2\varphi}=\cos\varphi + i\sigma^2 \sin\varphi .
\label{OrtTra}
\end{alignat}
This transformation replaces the the Pauli matrices 
in the spin-orbit couplings into  
$g^{\rm t}(\varphi){\bm \sigma}g(\varphi)=
(\cos2\varphi\,\sigma^1-\sin2\varphi\,\sigma^3,\sigma^2,
\cos2\varphi\,\sigma^3+\sin2\varphi\,\sigma^1)$.
On the other hand, 
since Eq. (\ref{OrtTra}) is an orthogonal transformation, 
one-parameter family of transformed Hamiltonian,
denoted by $H^\varphi$ or ${\cal H}^\varphi(\theta)$ below,
is equivalent. The pfaffian (\ref{Pfa}) is also invariant.
Therefore, when we are interested in the bulk properties,
we can deal with any Hamiltonian $H^\varphi$.

So far we have discussed the bulk properties.
However, if we consider finite periodic systems like Eq. (\ref{TwiBouCon}), 
a family of the Hamiltonian $H^\varphi$  behaves as different models.
It follows from Eq. (\ref{TwiBouCon}) that
the gauge transformation (\ref{OrtTra}) is commutative with
$e^{i\theta_2}$, but {\it not} with $e^{i\theta_1\sigma^3}$.
This tells us that spin-dependent twisted boundary 
condition is not invariant under the gauge transformation 
(\ref{OrtTra}), and breaks the gauge-equivalence of the Hamiltonian 
$H^\varphi$ which the bulk systems should have.

To understand this, the following alternative consideration  
may be useful:
If we want to study bulk properties of ${\cal T}$ invariant systems, 
we can start with any of $H^\varphi$. 
For one $H^\varphi$ with $\varphi$ fixed, let us impose the twisted
boundary condition (\ref{TwiBouCon}). After that, we can make a gauge
transformation (\ref{OrtTra}) back to $H^{0}$.
Then, we can deal with the original Hamiltonian $H^{0}$, but with a
gauge-dependent twisted boundary condition for $x$-direction;
\begin{alignat}{1}
&c_{j+L_1\hat 1}=
e^{i\theta_1 (\cos2\varphi\sigma^3-\sin2\varphi\sigma^1)}c_j.
\label{ModTwiBouCon}
\end{alignat}
Namely, the gauge equivalence is broken only by the boundary condition
in $1$-direction. 
Now, imagine a situation that at $\varphi=0$ the pfaffian (\ref{Pfa}) 
has one ${\cal T}$ pair of zeros. 
We denote them as 
$(\theta^*_1,\pm\theta_2^*)$ with vorticity $\pm m$.
Let us change $\varphi$ smoothly from 0 to $\pi/2$.
Then, it follows from Eq. (\ref{ModTwiBouCon}) that 
at $\varphi=\pi/2$ the coordinate of the torus is changed from 
$(\theta_1,\theta_2)$ into $(-\theta_1,\theta_2)$ and therefore,
we find the zeros at $(-\theta_1^*,\pm\theta_2^*)$ with vorticity $\mp m$.
We thus have a mapping 
$(\theta_1^*,\pm\theta_2^*)\rightarrow(-\theta_1^*,\mp\theta_2^*)$
for $\varphi=0\rightarrow \pi/2$.
Namely, the zero in the $\theta_2>0$ $(<0)$ space moves into the 
$\theta_2<0$ $(>0)$ space, and thus the zeros can move in the whole twist
space like those of the KM pfaffian.
During the mapping, there should occur a topological change, 
but it is attributed to the boundary, i.e, an artifact of broken
translational invariance, and
half of the zeros of twist pfaffian should be also classified by Z$_2$,
by taking into account the gauge transformation (\ref{OrtTra}).

Using this pfaffian,
we next show that its zeros can be
detectable by computing the spin Chern number which seems much easier to
perform than
searching them directly. 
The spin Chern number \cite{SWSH06} is defined by
$c_{\rm s}=\frac{1}{2\pi i}\int d^2\theta F_{12}(\theta)$,
where 
$F_{12}(\theta)\equiv
\partial_{1} A_2(\theta)-\partial_{2} A_1(\theta)$ 
is the field strength due to the U(1) part of
the (non-Abelian) Berry potential \cite{Hat04},
$A_\mu(\theta)\equiv{\rm tr}
\langle n(\theta)|\partial_{\mu}|m(\theta) \rangle$
with
$\partial_\mu=\partial/\partial\theta_\mu$.
First, we will show the relationship between  
the zeros of the twist pfaffian  
and the spin Chern number $c_{\rm s}$.
For the time being, we fix $\varphi$.
The degenerate ground state as the $M$-dimensional multiplet has a local U$(M)$ gauge 
degree of freedom,
$|n(\theta)\rangle\rightarrow \sum_m|m(\theta)\rangle V_{mn}(\theta)$,
where $V(\theta)$ is a unitary matrix \cite{Hat04}.
Let us 
denote
$V(\theta)=e^{i\alpha(\theta)/M}\tilde V(\theta)$,
where $\det\tilde V(k)=1$.
This transformation induces 
$A_\mu(\theta)\rightarrow A_\mu(\theta)
+i\partial_{\mu}\alpha(\theta)$
to the Berry potential.
If one can make gauge-fixing globally over the whole twist space,
the Chern number is proved to be zero. Only if the global gauge-fixing 
is impossible, the Chern number can be nonzero.

Among various kinds of gauge-fixing, we can use the gauge that 
{\it $p(\theta)$ is real positive}, because 
$p(\theta)\rightarrow p(\theta) \det V(\theta)=p(\theta)e^{i\alpha(\theta)}$.
This rule can fix the gauge of the Berry potential except for $p(\theta)=0$.
Therefore, nontrivial spin Chern number is due to an obstruction to the 
smooth gauge-fixing by the twist pfaffian.
This correspondence also proves that 
{\it the spin Chern number is an even integer}, since 
the pfaffian (\ref{Pfa}) always has ${\cal T}$ pair zeros, 
and since $F_{12}(\theta_1,-\theta_2)=F_{12}(\theta_1,\theta_2)$. 

Let us now 
change  $\varphi$.
At $\varphi=0$, we obtain a certain integral $c_{\rm s}$.
Remember that at $\varphi=\pi/2$, the coordinate of the torus 
is changed into $(-\theta_1,\theta_2)$.
Therefore, we have a mapping $c_{\rm s}\rightarrow -c_{\rm s}$
for $(\theta_1,\theta_2)\rightarrow (-\theta_1,\theta_2)$.
As in the case of the pfaffian (\ref{Pfa}), we expect 
topological changes during the mapping. 
However, as stressed, these changes are accompanied by
{\it no gap-closing in the bulk spectrum}, 
just induced by the symmetry breaking 
{\it boundary} term which is an artifact to define the Chern numbers. 
Therefore, the states with $\pm c_{\rm s}$ should 
belong to an equivalent topological sector.
Since the minimum nonzero spin Chern number is 2, we expect
$c_{\rm s}$ mod 4 
(if we define the spin Chern number in half the
twist space, mod 2) classifies the topological sectors.

So far we have discussed the Z$_2$ characteristics of the spin Chern
number $c_{\rm s}$. We next present several examples.
To this end, we employ an efficient method of
computing Chern numbers proposed in Ref. \cite{FHS05}.
We first discretize the twist space
$[0,2\pi]\otimes [0,2\pi]$ into a 
square lattice such that $\theta_\mu=2\pi j_\mu/N_\mu$,
where $j_\mu=1,\cdots,N_\mu$ \cite{FooNot2}. 
We denote the sites on this lattice as $\theta_\ell$ with
$\ell=1,2,\cdots, N_1N_2$. 
We next define a U(1) link variable
associated with the ground state multiplet of the dimension $M$,
\begin{eqnarray*}
U_\mu(\theta_\ell)=|\det{\bm U}_\mu(\theta_\ell)|^{-1}
\det {\bm U}_\mu(\theta_\ell) ,
\end{eqnarray*}
where
${\bm U}_\mu(\theta_\ell)_{mn}=
\langle m(\theta_\ell)|n(\theta_\ell+\hat\mu)\rangle$ 
with $n,m=1,\cdots,M$ denotes the (non-Abelian) Berry link variable.
Here, $\hat\mu$ is the vector in $\mu$ direction with 
$|\hat\mu|=2\pi/N_\mu$.
Next define the lattice field strength,
\begin{eqnarray*}
F_{12}(\theta_\ell)=
\ln U_1(\theta_\ell)U_2(\theta_\ell+\hat 1)
U^{-1}_1(\theta_\ell+\hat 2)U^{-1}_2(\theta_\ell) ,
\end{eqnarray*}
where we choose the branch of the logarithm as $|F_{12}(\theta_\ell)|<\pi$.
Finally, manifestly {\it gauge invariant} lattice Chern number is obtained:
\begin{eqnarray}
c_{\rm s}=\frac{1}{2\pi i}\sum_\ell F_{12}(\theta_\ell) .
\label{LatCheNum}
\end{eqnarray}
As shown in \cite{FHS05}, the spin Chern number thus defined is strictly
{\it integral}. 
To see this, let us introduce a lattice gauge potential
$A_\mu(\theta_\ell)=\ln U_\mu(\theta_\ell)$
which is also defined in $|A_\mu(\theta_\ell)|<\pi$.
Note that this field is periodic, 
$A_\mu(\theta_\ell+N_\mu)=A_\mu(\theta_\ell)$.
Then, we readily find
\begin{eqnarray}
F_{12}(\theta_\ell)=\Delta_1 A_2(\theta_\ell)-\Delta_2A_1(\theta_\ell)+
2\pi i n_{12}(\theta_\ell) ,
\label{FieStrN}
\end{eqnarray}
where $\Delta_\mu$ stands for the difference operator and 
$n_{12}(\theta_\ell)$ is a local {\it integral }field which is
referred to as $n$-field.
Finally, we reach
$c_{\rm s}=\sum_\ell n_{12}(\theta_\ell)$.
This completes the proof that the spin Chern number is integral.
While the $n$-field depends on a gauge we adopt, the sum is invariant. 
For the ${\cal T}$ invariant systems, 
{\it the pfaffian
(\ref{Pfa}) is very useful for the gauge-fixing} also for the lattice
computation. 
In the continuum theory, we have stressed that the zeros of the pfaffian
play a central role in the Z$_2$ classification. Since such zeros occur at 
several specific points in the twist space, it is very hard
to search them numerically.

Contrary to this feature in the continuum approach, 
we can detect the zeros in the lattice approach as follows:
Suppose that we obtain an exact Chern
number using Eq. (\ref{LatCheNum}) with sufficiently large $N_\mu$. 
Since the lattice Chern number is topological (integral), 
which implies that even if we slightly change the lattice
(e.g, change the lattice size or infinitesimally shift
the lattice), 
the Chern number remains unchanged.
Next,  let us compute the $n$-field in the gauge that
{\it the pfaffian is real positive}.
If the zeros of the pfaffian happen to locate 
on sites of the present lattice, we cannot make gauge-fixing.
Even in such cases, we can always avoid the zeros of the pfaffian 
by redefining the lattice with the spin Chern number kept unchanged.
Thus we can always 
compute the well-defined $n$-field configuration. 
If the Chern number is nonzero, there exist nonzero $n$-field
somewhere in the twist space. 
These nonzero $n$-field occur in general near the zeros of the
pfaffian: Thus, without searching zeros of pfaffian in the continuum
twist space, we can observe {\it 
the trace of such zeros as nonzero $n$-fields}.

Now let us study a graphene model with spin-orbit couplings
\cite{KanMel05a,KanMel05b,SWSH06};
\begin{alignat}{1}
H=&-t\sum_{\langle i,j\rangle}c^\dagger_ic_j
+\frac{2i}{\sqrt{3}}V_{\rm so}
\sum_{\langle\langle i,j\rangle\rangle}c^\dagger_i{\bm \sigma}\cdot
\left({\bm d}_{kj}\times{\bm d}_{ik}\right)c_j 
\nonumber\\& 
+iV_{\rm R}\sum_{\langle i,j\rangle}
c^\dagger_i\left({\bm\sigma}\times{\bm d}_{ij}\right)^3c_j
+v_{\rm s}\sum_j\mbox{sgn} j~c^\dagger_jc_j,
\label{Ham} 
\end{alignat}
where
$c^\dagger_j=(c^\dagger_{\uparrow j}, c^\dagger_{\downarrow j})$
is the electron creation operator at site $j$ on the honeycomb lattice,
$\mbox{sgn}j$ denotes $1~(-1)$ if $j$ belongs to the A (B) sublattice,
and ${\bm d}_{ij}$ stands for the vector from $j$ to $i$ sites.
The first term is the nearest neighbor hopping,
the second term is the $s_z$-conserving 
next-nearest neighbor spin-orbit coupling, whereas the third term is
the Rashba spin-orbit coupling, and final term is a 
on-site staggered potential. 
Analyzing the KM pfaffian,
Kane and Mele \cite{KanMel05a} have derived the phase diagram:
It is in the QSH phase for 
$|6\sqrt{3}V_{\rm so}-v_{\rm s}|>\sqrt{v_{\rm s}^2+9V_{\rm R}^2}$ 
and in the insulating phase otherwise.
Sheng {\it et. al.} \cite{SWSH06} have computed the spin Chern
number $c_{\rm s}=2$ in the QSH phase and 0 in the insulating phase.

\begin{figure}[htb]
\begin{tabular}{ccc}
\includegraphics[width=.33\linewidth]{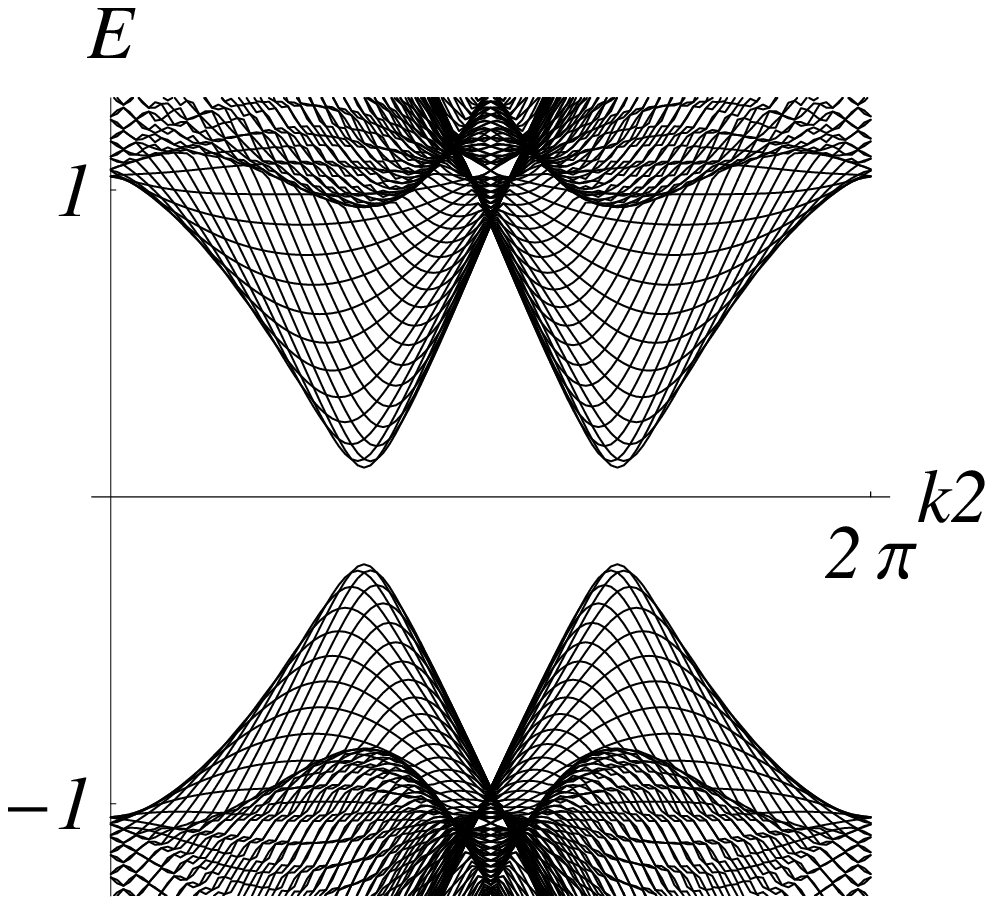}
&\includegraphics[width=.33\linewidth]{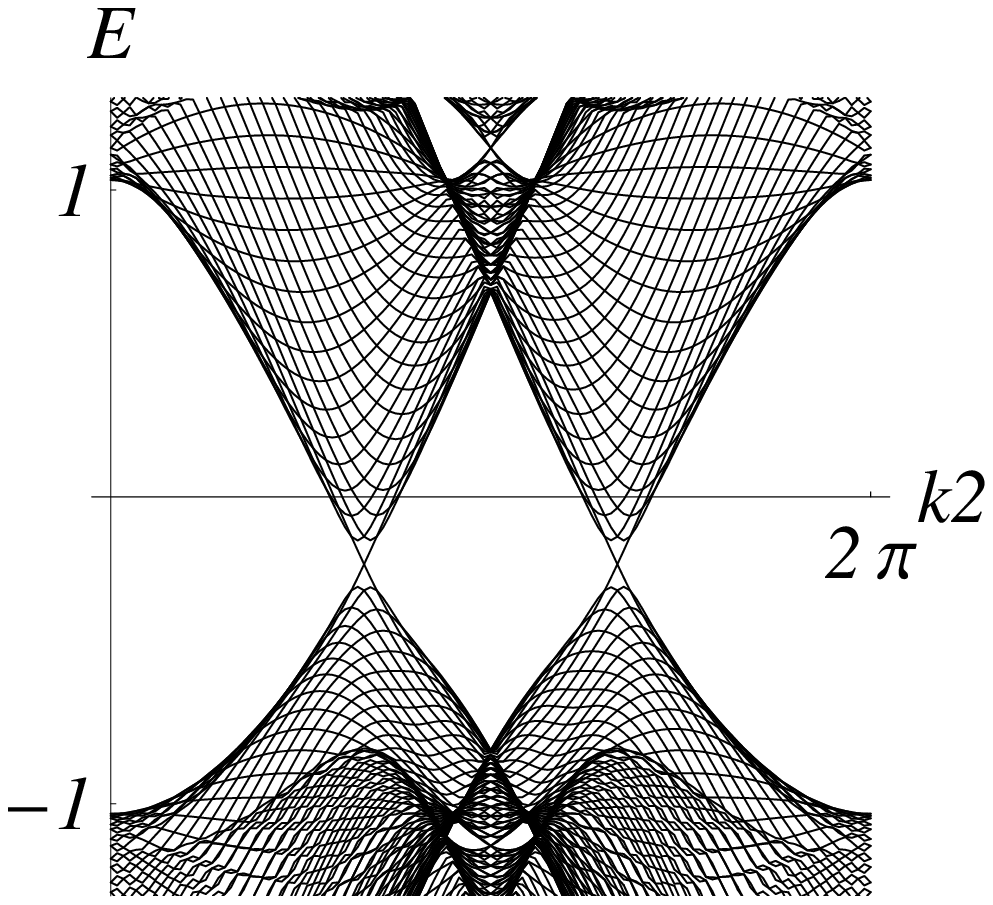}
&\includegraphics[width=.33\linewidth]{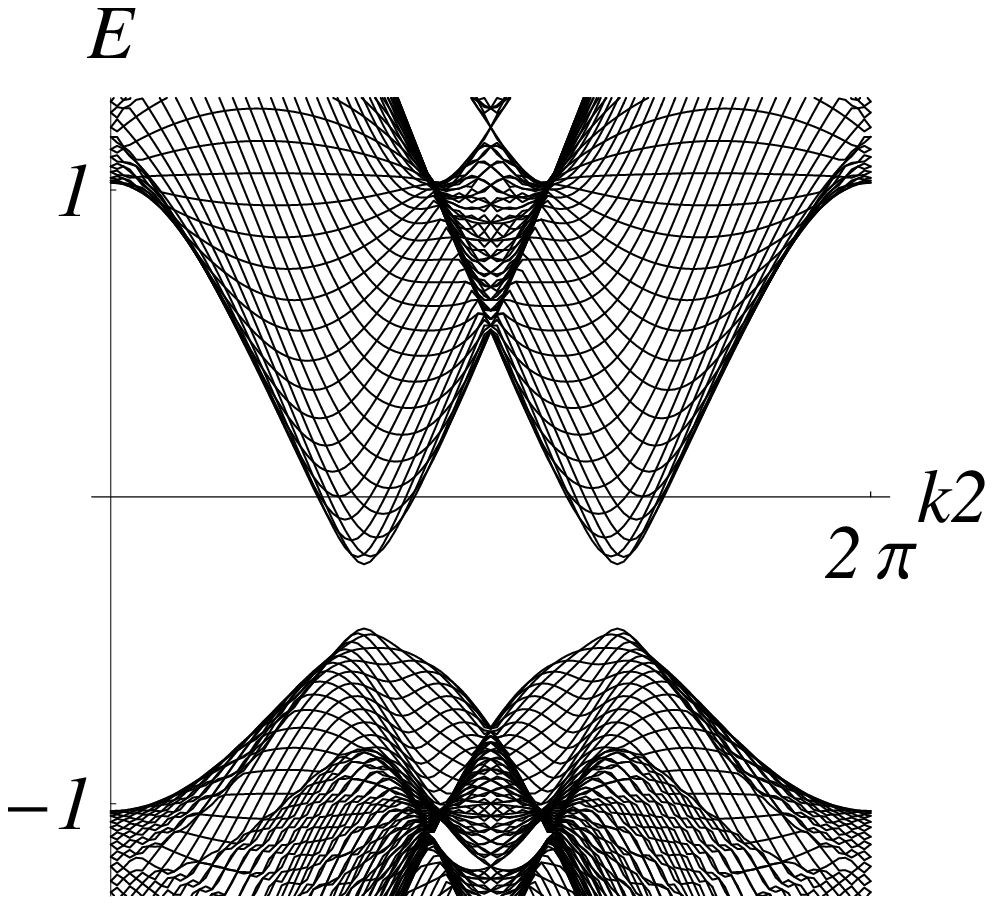}
\end{tabular}
\caption{Spectrum for $\theta_1=0$ as a function of $k_2$. 
Parameters used are $V_{\rm so}=0.1t$, $v_{\rm s}=0.3t$, 
and $V_{\rm R}=0.1t$ (left), $V_{\rm R}=0.225207t$ (middle),
and $V_{\rm R}=0.3t$ (right). 
}
\label{f:Spectrum}
\end{figure}

For numerical computations, it is convenient to use the momentum $k_2$ 
\cite{FooNot1,SWSH06}
instead of $\theta_2$ because of the translational invariance along this
direction even with the boundary condition (\ref{TwiBouCon}).  
First, we show the spectrum in Fig. \ref{f:Spectrum} at $\theta_1=0$ as
a function of $k_2$.
The left belongs to the QSH phase with $c_{\rm s}=2$, whereas
the right to the insulating phase with $c_{\rm s}=0$.
This topological change is due to the gap-closing in the {\it bulk}
spectrum, as shown in the middle in Fig. \ref{f:Spectrum}. 
Therefore, the phase with $c_{\rm s}=2$ is topologically 
distinguishable from the phase with $c_{\rm s}=0$.

\begin{figure}[htb]
\begin{tabular}{ccc}
\includegraphics[width=.33\linewidth]{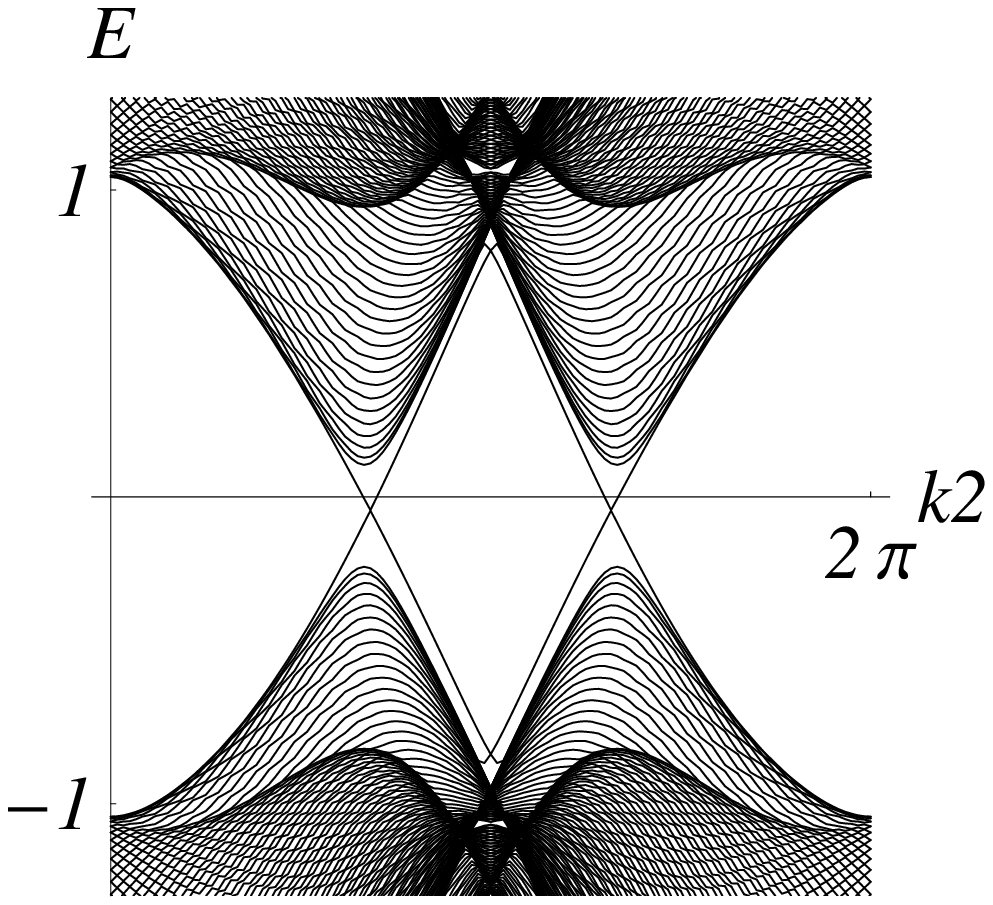}
&\includegraphics[width=.33\linewidth]{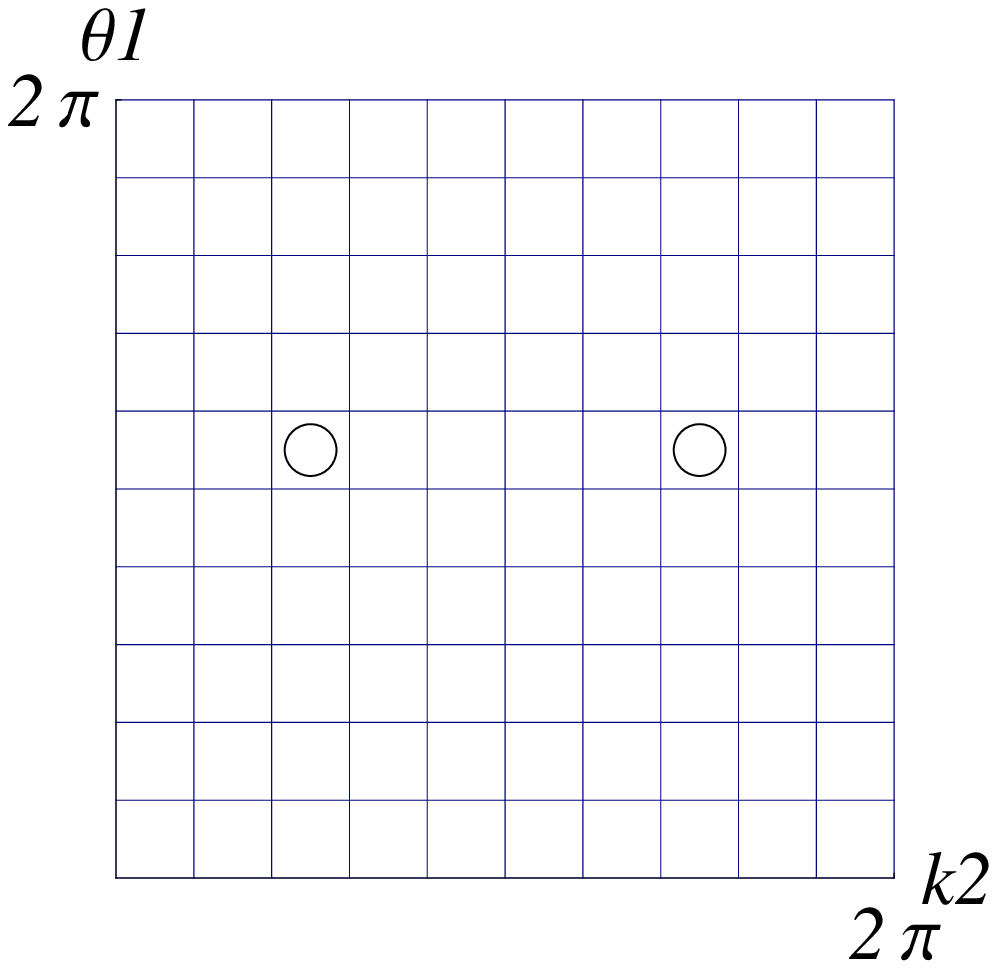}
\hspace{-3mm}
&\includegraphics[width=.33\linewidth]{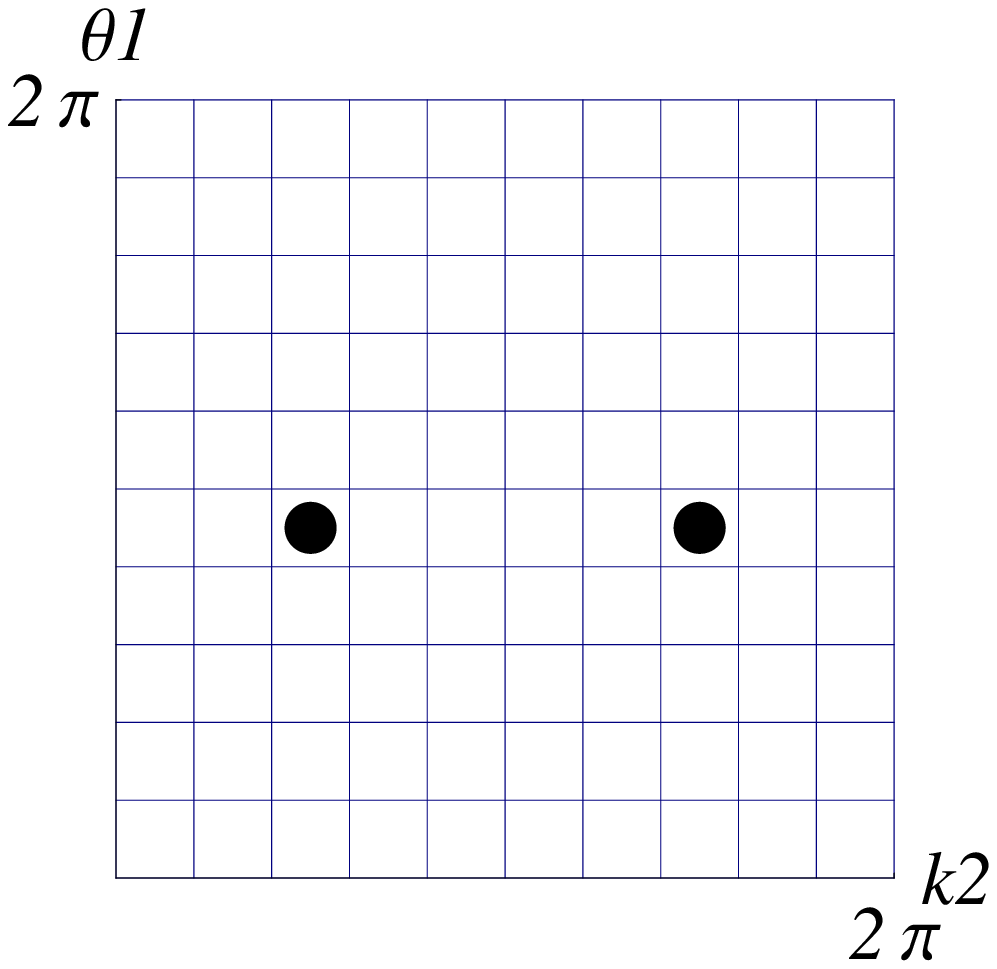}
\end{tabular}
\caption{Left: Spectrum  
for nonzero $\varphi=\pi/4$ as a function of 
$k_2$ at $\theta_1=\pi/2$.
Other parameters are the same as those of the left 
in Fig. \ref{f:Spectrum}.
 Right two: 
the $n$-field configuration corresponding to the left in
 Fig. \ref{f:Spectrum}. The left (right) is at $\varphi=0~(\pi/2)$.
The white (black) circle denotes $n=1$ $(-1)$, 
while the blank means $n=0$. 
We used the meshes $N_1=N_2=10$.
}
\label{f:nfield}
\end{figure}

Contrary to this, how about the phases with $c_{\rm s}=\pm2$?
As we have mentioned, the phase with $c_{\rm s}$ changes 
into $-c_{\rm s}$ when 
we vary $\varphi$ from 0 to $\pi/2$. In this process,
{\it boundary}-induced topological changes must occur. 
We show in Fig. \ref{f:nfield} the spectrum cut at $\theta_1=\pi/2$
for $\varphi=\pi/4$.
We indeed observe a gap-closing at finite $\theta_1$,
and the spin Chern number $c_{\rm s}=2$ for $0\leq\varphi<\pi/4$
is changed into $c_{\rm s}=-2$ for $\pi/4<\varphi\leq \pi/2$.
As stressed, 
this change is attributed to the boundary (edge states in Fig. \ref{f:nfield}), 
not to the bulk, and we conclude that
the phase $c_{\rm s}=\pm2$ is classified as the same QSH phase.
These spin Chern numbers $c_{\rm s}=\pm2$ are well visible by the $n$-field.
In Fig. \ref{f:nfield}, we also show the $n$-field 
for $\varphi=0$ and $\pi/2$ cases in the QSH phase. The points of nonzero
$n$-field is closely related with the positions of the pfaffian zeros. 
We also note that in Eq. (\ref{LatCheNum})
net contributions to the
nonzero Chern number is just from two points. 

Next, let us study a bilayer graphene.
Suppose that we have  two decoupled
sheets of graphene described by $H^{\varphi_i}$ with $i=1,2$ whose lattices
include $A$, $B$ sites and $\tilde A$, $\tilde B$ sites, respectively. 
For simplicity, we take into account only the interlayer coupling $\gamma_1$
between $\tilde A$ and $B$ \cite{McCFal06}:
$V_{12}=\gamma_1\sum_{j}c_{1\tilde A,j}^\dagger c_{2B,j}+\mbox{h.c.}$,
where $i=1,2$ in $c_{i,j}$ indicate the $i$th sheet.
Now make  the gauge transformation (\ref{OrtTra}) separately for
each sheet to obtain the same $H^0$ as Eq. (\ref{Ham}). 
Then, we have identical bilayer system $H^0\otimes H^0$ coupled by
$V_{12}=\gamma_1\sum_{j}
c_{1\tilde A,j}^\dagger g(\varphi_1) 
g^{\rm t}(\varphi_2)c_{2B,j+}\mbox{h.c.}$
with two independent boundary conditions
$c_{i,j+L_1\hat 1}=
e^{i\theta_1 (\cos2\varphi_i\sigma^3+\sin2\varphi_i\sigma^1)}c_{i,j}$.
\begin{figure}[htb]
\begin{tabular}{ccc}
\includegraphics[width=.33\linewidth]{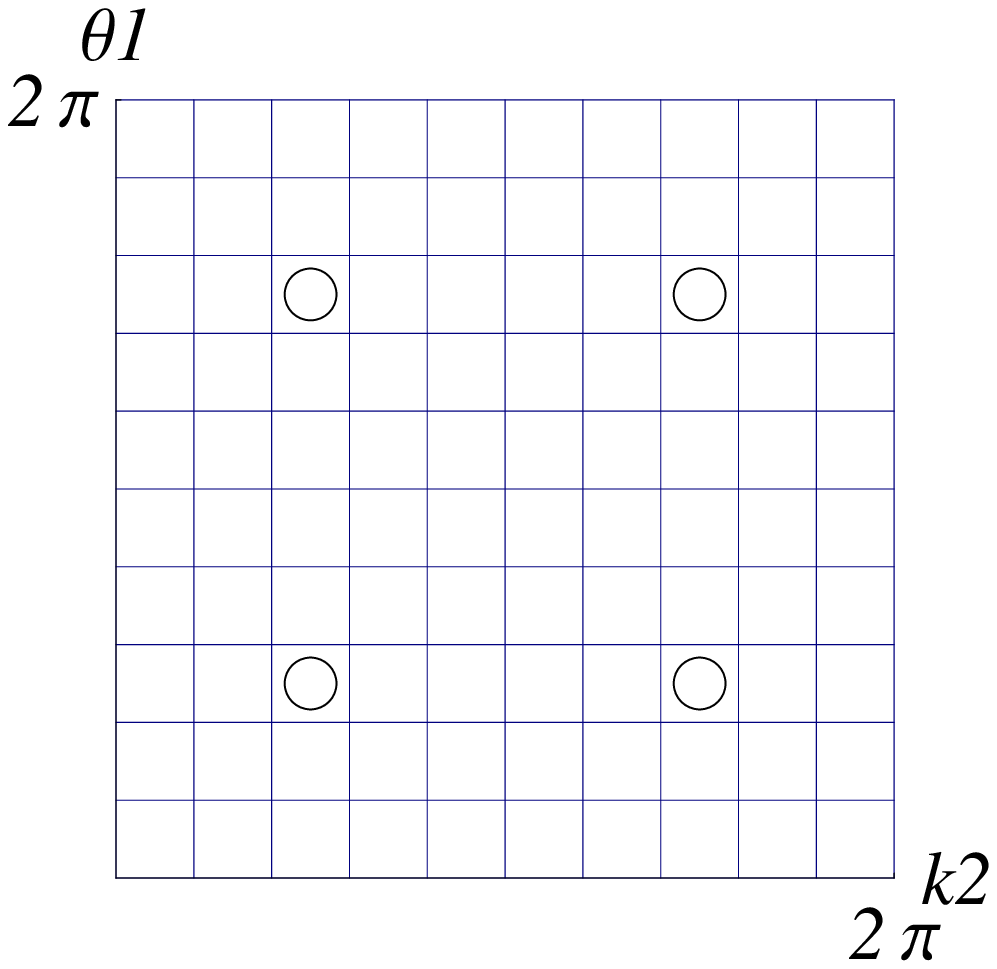}
\hspace{-3mm}
&\includegraphics[width=.33\linewidth]{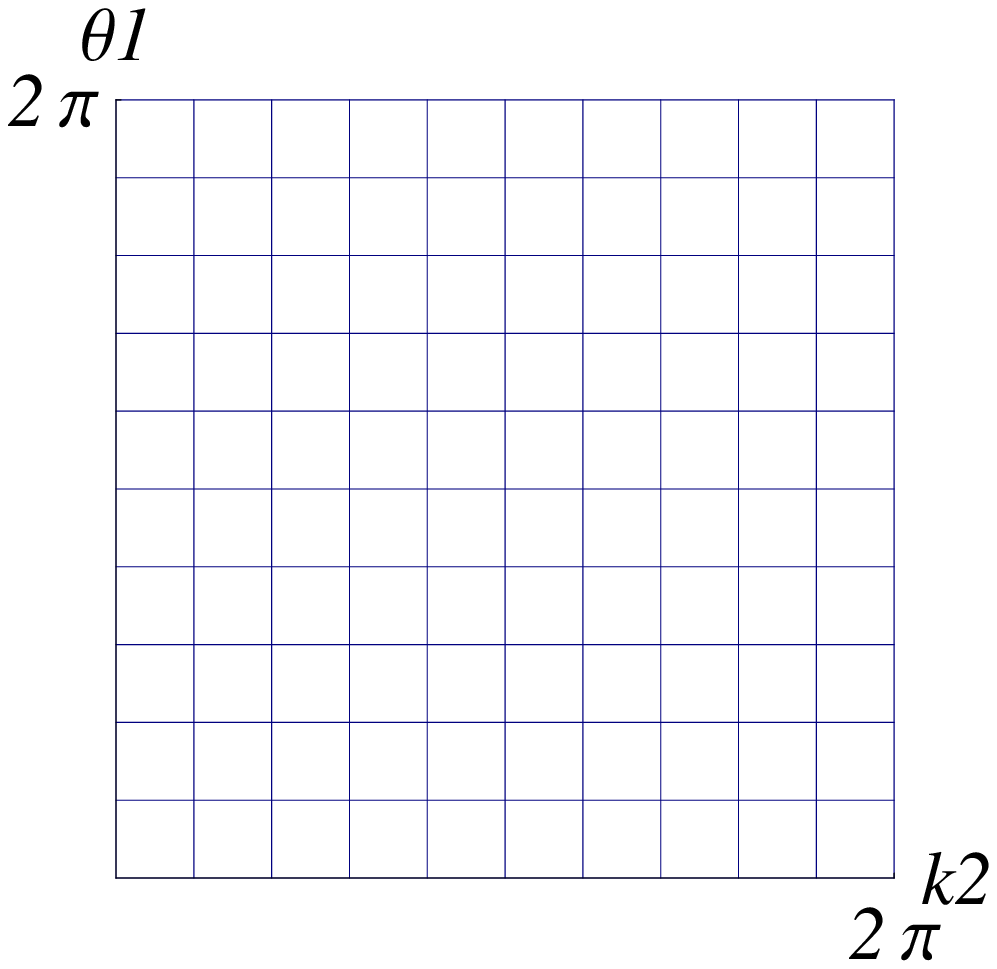}
\hspace{-3mm}
&\includegraphics[width=.33\linewidth]{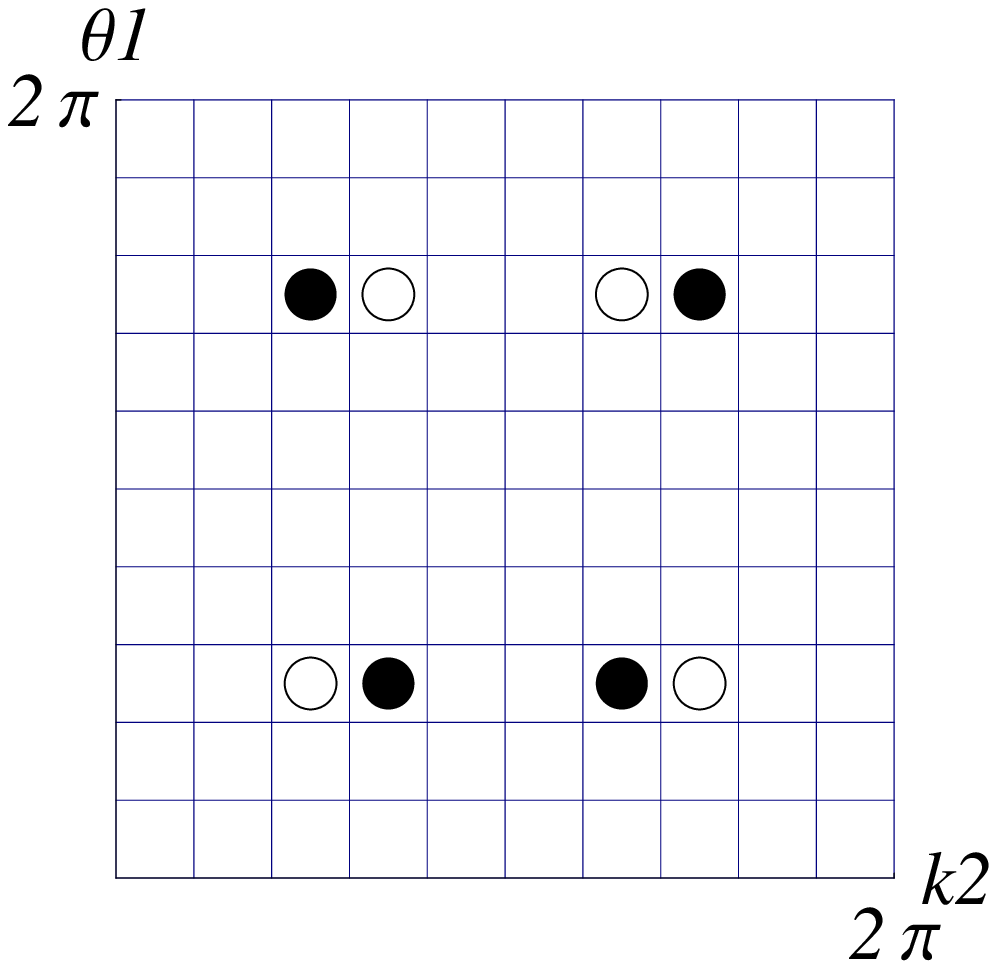}
\end{tabular}
\caption{The $n$-field configuration for $\gamma_1=0.1t$.
Other parameters are same as those of the left in Fig. \ref{f:Spectrum}.
Left: $\varphi_1=0$ and $\varphi_2=0$.
Middle: $\varphi_1=\pi/2$ and $\varphi_2=0$.
Right: $\varphi_1=\pi/4$ and $\varphi_2=-\pi/4$.
In the case $\varphi_1=\pi/2$ and $\varphi_2=\pi/2$ we have the same
figure as the left but with black circles.
We used the meshes 
$N_1=N_2=10$. 
}
\label{f:NfieldDL}
\end{figure}

For the same parameters as those of the left in Fig. \ref{f:Spectrum}, 
the spin Chern number is, of course, $c_{\rm s}=2+2=4$ in the limit
$\gamma_1=0$.
This spin Chern number remains unchanged for small but finite interlayer
coupling $\gamma_1$. 
However, taking into account the gauge transformation $g(\varphi_i)$,
the spin Chern number changes. 
In Fig. \ref{f:NfieldDL}, we show examples of the $n$-field for $\gamma_1=0.1t$.
We have the spin Chern numbers $2+2=4$, $2-2=0$, and $-2-2=-4$:
All of them are denoted as $c_{\rm s}=0$ mod 4, which belong to
the insulating phase.
Detail analysis of this model including the interlayer coupling $\gamma_3$ 
will be published elsewhere.

Finally, 
we comment that the QSH effect is understood by the edge states 
\cite{KanMel05b}, and therefore, it is interesting to establish the 
bulk-edge correspondence \cite{Hat93} for ${\cal T}$ invariant systems
with respect to Z$_2$.
We also mention that 
Fu and Kane \cite{FuKan06} and Moore and Balents \cite{MooBal06}
have recently discussed the relationship between the Z$_2$
order and the spin Chern number, and reached a similar conclusion.

This work was supported in part 
by Grant-in-Aid for Scientific Research  
(Grant No. 17540347, No. 18540365) from JSPS
and on Priority Areas (Grant No.18043007) from MEXT.


\begin{thebibliography}{99}
\bibitem{FooNot2}
We have shifted $\theta_\mu$, for numerical conveniences.
\bibitem{FooNot1}
In this case, we 
use discrete $k_2$ instead of $\theta_2$.
Then, $L_2=N_2$ 
means the number of the meshes for discrete $k_2$.
\end{thebibliography}
\end{document}